%% file: article.tex
\documentclass[journal=jacsat,manuscript=article]{achemso}
\SectionNumbersOn

\usepackage{chemformula} % Formula subscripts using \ch{}
\usepackage[T1]{fontenc} % Use modern font encodings
 \usepackage[version=3]{mhchem} % Formula subscripts using \ce{}
  \usepackage{booktabs}  
  \usepackage{threeparttable}  
  \usepackage{multirow}
  
  \usepackage[numbers,sort&compress]{natbib}
  \usepackage{amsmath}
  \usepackage{amsfonts}
  \usepackage{longtable,booktabs}

\input{sections/authors.tex}

\title[DeepScaffold]
{
    DeepScaffold:
    a comprehensive tool for scaffold-based de novo drug discovery using deep 
    learning
}


\begin{document}

\input{sections/abstract.tex}

\input{sections/introduction.tex}
\input{sections/methods.tex}
\input{sections/results.tex}

\input{sections/conclusion.tex}

\bibliography{article}

\end{document}

%% file: sections/authors.tex
\author{Yibo Li}
\affiliation{
    State Key Laboratory of Natural and Biomimetic Drugs, 
    School of Pharmaceutical Sciences, 
    Peking University, Xueyuan Road 38, Haidian District, 
    100191, Beijing, China
}
\alsoaffiliation{Stonewise, Beijing, China}
\altaffiliation{Contributed equally to this work}

\author{Jianxing Hu}
\affiliation{
    State Key Laboratory of Natural and Biomimetic Drugs, 
    School of Pharmaceutical Sciences, 
    Peking University, Xueyuan Road 38, Haidian District, 
    100191, Beijing, China
}
\altaffiliation{Contributed equally to this work}

\author{Yanxing Wang}
\affiliation{
    State Key Laboratory of Natural and Biomimetic Drugs, 
    School of Pharmaceutical Sciences, 
    Peking University, Xueyuan Road 38, Haidian District, 
    100191, Beijing, China
}

\author{Jielong Zhou}
\email{zhoujielong@stonewise.cn}
\affiliation{Stonewise, Beijing, China}

\author{Liangren Zhang}
\email{liangren@bjmu.edu.cn}
\affiliation{
    State Key Laboratory of Natural and Biomimetic Drugs, 
    School of Pharmaceutical Sciences, 
    Peking University, Xueyuan Road 38, Haidian District, 
    100191, Beijing, China
}

\author{Zhenming Liu}
\email{zmliu@bjmu.edu.cn}
\affiliation{
    State Key Laboratory of Natural and Biomimetic Drugs, 
    School of Pharmaceutical Sciences, 
    Peking University, Xueyuan Road 38, Haidian District, 
    100191, Beijing, China
}

%% file: sections/abstract.tex
\begin{abstract}
  The ultimate goal of drug design is to find novel compounds with desirable 
  pharmacological properties.  Designing molecules retaining  particular 
  scaffolds as the core structures of the molecules is one of the  efficient 
  ways to obtain potential drug candidates with desirable properties. We 
  proposed a scaffold-based molecular generative model for scaffold-based drug 
  discovery, which performs molecule generation based on a wide spectrum of 
  scaffold definitions, including BM-scaffolds, cyclic skeletons, as well as 
  scaffolds with specifications on side-chain properties. The model can 
  generalize the learned chemical rules of adding atoms and bonds to a given 
  scaffold. Furthermore, the generated compounds were evaluated by molecular 
  docking in DRD2 targets and the results demonstrated that this approach can 
  be effectively applied to solve several drug design problems, including the 
  generation of compounds containing a given scaffold and de novo drug design of 
  potential drug candidates with specific docking scores.
\end{abstract}

%% file: sections/introduction.tex
\section{Introduction}

"Molecular scaffold" is one of the most important and widely used concepts in 
medicinal chemistry. Given a chemical compound, its "scaffold" generally 
represents the core structures of the molecular framework \cite{RN242}. 
Although various definitions of molecular scaffolds are available, 
the most widely adopted version is given by Bemis and Murcko \cite{RN13}, 
who obtain the scaffold by removing all side chains (or R-groups). 
Among all scaffolds, those with preferable bioactivity properties 
(which are referred to as "privileged scaffolds"\cite{RN14} or 
"bioactive scaffolds"\cite{RN16,RN248}) are of particular interest to drug 
discovery. Privileged scaffolds are often used as a starting point for 
compound synthesis or diversification. Previous researches have developed 
various types of tools to extract and utilize information from scaffolds, 
including tools for the discovery of privileged scaffolds (such as CSE 
\cite{RN259} and \cite{RN248}), tools for organizing scaffolds (such as 
HierS \cite{RN247}, SCONP \cite{RN241}, ST\cite{RN15} and SN\cite{RN16}), 
as well as tools for visualization (such as Scaffold Hunter \cite{RN254}).

Recently, deep learning \cite{RN158} have revealed
itself as a new and promising tool for many drug-design-related
problems. In particular, there is a growing interest in using deep
generative models\cite{RN177} for \emph{de novo}
molecule design\cite{RN43}. Early works in this area
\cite{RN191,RN234,RN44} mainly utilizes RNN based
architectures (such as GRU \cite{RN44} or LSTM
\cite{RN191}) to generate SMILES strings
\cite{RN129}, which is a serialized representation of
molecules graph. Those methods have proven to be very effective for
\emph{de novo} molecule design, but still faces certain issues, such as
the need to learn SMILES gramma. Recent works
\cite{RN236,RN48,RN49,RN229,RN228} have improved
upon this approach by using graph-based representations of molecules.
Also, other types of generative architecture have been explored,
such as GAN \cite{RN251}. Those models have been
successfully applied to various drug design problems, such as
the property-based design of virtual libraries \cite{RN236,RN44},
ligand\cite{RN191,RN236,RN44} and structural\cite{RN257} based drug design, 
as well as local optimization of molecular structures\cite{RN44,RN228}.

With the promising performance, it is natural to apply deep generative
models to scaffold directed drug design. Early works have used
conditional generative model \cite{RN236},
reinforcement learning \cite{RN236} or transfer
learning \cite{RN341} to achieve molecule generation
based on scaffold queries. However, those methods do not guarantee that
the outputs molecule will match the input query, and are difficult to be
extended to new scaffolds. A recent publication by Lim et. al.
\cite{RN339} solved the problems above using a model
that directly grow molecules on the given scaffold. However, the model
is restricted only to BM-scaffolds, which does not necessarily fit well
with all tasks in medicinal chemistry.

In this work, we introduce DeepScaffold, a novel and comprehensive
solution for scaffold-based drug discovery. Different from previous
methods, DeepScaffold can perform molecule generation based on a wide
spectrum of scaffold definitions, including BM-scaffolds, cyclic
skeletons, as well as scaffolds with specifications on side-chain
properties. Our proposed method can guarantee that the output molecule
would match the scaffold query, and can be easily extended to new
scaffolds outside the training set. A broad set of evaluation is carried
out to access the performance of DeepScaffold. We also discuss several
challenges faced when evaluating such scaffold-based generative models.

%% file: sections/methods.tex
\section{Methods}
\input{sections/methods/overview.tex}
\input{sections/methods/data.tex}
\input{sections/methods/csk.tex}
\input{sections/methods/scaffold.tex}
\input{sections/methods/pharmacophore.tex}
\input{sections/methods/evaluation.tex}

%% file: sections/methods/overview.tex
\begin{figure}[t!]
    \centering
    \includegraphics{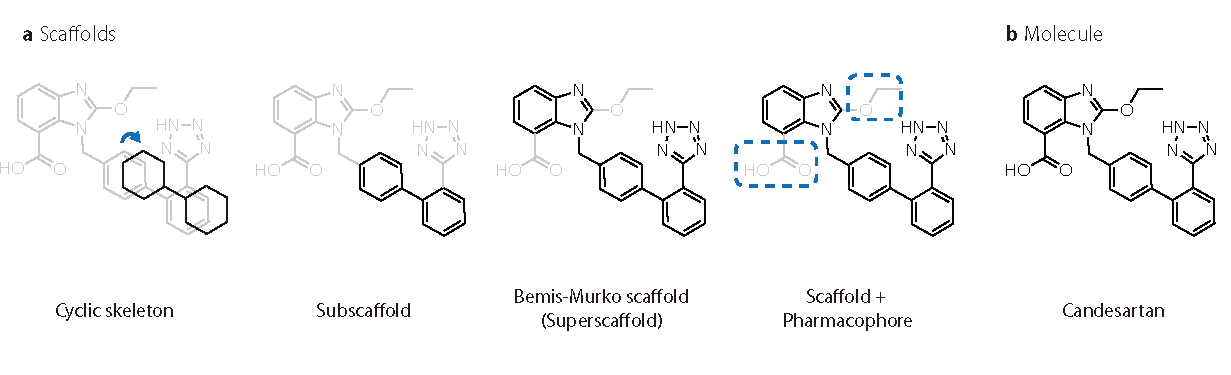}
    \caption{
        An overview of DeepScaffold. DeepScaffold
        is a comprehensive tool for scaffold-directed drug discovery, capable of
        performing molecule generation based on \textbf{a.} cyclic skeletons
        (CSKs), \textbf{b.} classical molecular scaffolds, such as Bemis-Murko
        scaffolds, as well as \textbf{c.} scaffolds with additional
        pharmacophore-based queries for side chains.
    }
    \label{fig:overview-1}
\end{figure}

\subsection{An overview of DeepScaffold}
\label{sec:method-overview}

Before diving into the implementation details, we first take a look at
the overall architecture of DeepScaffold. As mentioned earlier,
DeepScaffold is capable of performing de novo drug design based on the
following scaffold definitions (Figure\ref{fig:overview-1}):

\begin{itemize}
    \item
    Cyclic skeletons
    \item
    Scaffolds with atom and bond type information, such as Bemis-Murko
    scaffold
    \item
    Scaffolds with additional specification on side-chain properties
\end{itemize}

\begin{figure}[t!]
    \centering
    \includegraphics{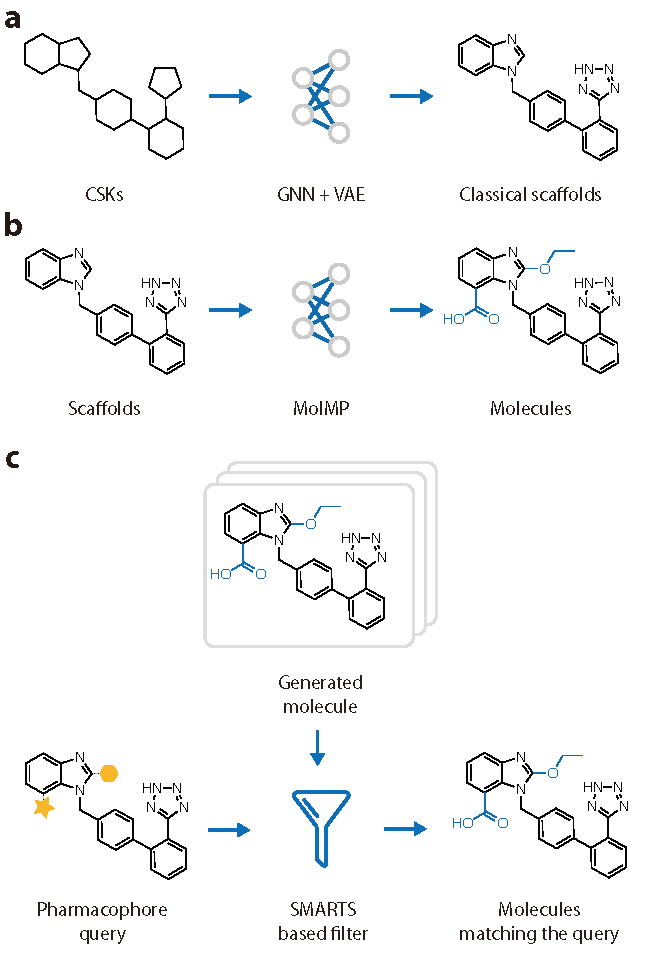}
    \caption{
        The basic components for DeepScaffold. As
        shown in the figure, DeepScaffold is composed of the following three
        models: \textbf{a.} A neural network for completing atom and bond type
        for cyclic skeletons; \textbf{b.} A scaffold-based molecule generative
        model; \textbf{c.} A filter based on side-chain properties
    }
    \label{fig:overview-2}
\end{figure}

For convenience, we refer to the first type of structures as the
``skeletons'', and the second type of structure as the ``classical
scaffolds''. To achieve those functionalities, DeepScaffold is composed
of the following three components:

\begin{itemize}
    \item
    A neural network for completing atom and bond type for cyclic
    skeletons (Figure \ref{fig:overview-2}\textbf{a})
    \item
    A scaffold-based molecule generator (Figure \ref{fig:overview-2}\textbf{b})
    \item
    A filter for side-chain properties based on pharmacophore queries 
    (Figure \ref{fig:overview-2}\textbf{c})
\end{itemize}

The three components are discussed in detail in the rest part of this
section. We first discuss the dataset and data processing workflow in
Section \ref{sec:dataset}.The transformer from skeletons to scaffolds is discussed
in Section \ref{sec:method-csk}, and the scaffold-based generator is discussed in
Section \ref{sec:method-scaffold}. We briefly discuss the pharmacophore quires in \ref{sec:method-pharmacophore}
Finally, the evaluation methods are discussed in
Section \ref{sec:eval}.

%% file: sections/methods/data.tex
\subsection{Dataset}
\label{sec:dataset}

\begin{figure}[t!]
    \centering
    \includegraphics{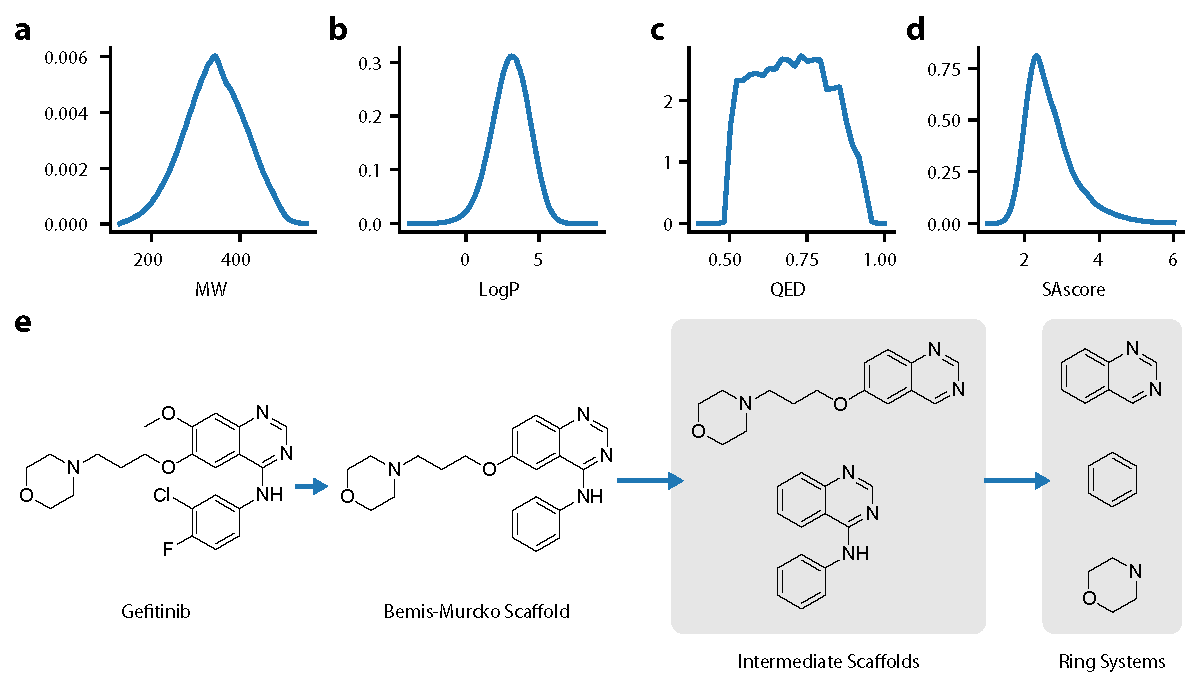}
    \caption{
        a-d. The distribution of several important molecular
        properties (a for MW, b for LogP, c for QED, d for SAscore) of the
        extracted molecules. e. Workflow for scaffold extraction. Given the
        input molecule, the Bemis-Murcko scaffold is first extracted by removing
        all side chains. Other scaffolds are then obtained by recursively
        removing ring systems.
    }
    \label{fig:data-processing}
\end{figure}

For model training, we use molecules from ChEMBL
\cite{RN24}, which is a widely used database containing
both structural and bioactivity information for molecules. The data
processing workflow largely follows that used in our previous study
\cite{RN236}. First of all, the structural information
of molecules represented as canonical SMILES are first extracted from
ChEMBL dataset. The molecules are then standardized using RDKit, which
involves the removal of salt and isotopes, as well as charge
neutralization. Molecules containing elements outside the set \{H, C, N,
O, F, P, S, Cl, Br, I\} are filtered from the dataset. Also, to
make the dataset more suitable for drug design purposes, we only keep
molecules with QED \cite{RN26} larger than 0.5. The
final dataset (denoted as \(\mathcal{M}=\{m_i\}_{i=1}^{N_m}\)) contains
\(N_m=914464\) molecules. The distribution of several important
molecular properties (including MW, LogP, QED and SAscore
\cite{RN20}) are visualized in Figure \ref{fig:data-processing}.

After the dataset is obtained, we need to extract scaffolds from each
molecule. Previous researchers have developed many different ways to
extract and organize scaffolds from structural data. Examples include
HierS \cite{RN247}, SCONP
\cite{RN241}, scaffold tree (ST)
\cite{RN15} and scaffold network (SN)
\cite{RN16}. Here, we adopt the same scaffold
extraction method as HierS. Compared with ST and SN, which are more
widely adopted, HierS results in a fewer amount scaffolds, but is much
easier to implement.

Molecules are composed of three components: ring systems (ring), side-chain 
bonds and atoms (chain), and linking bonds and atoms (linker). Atoms that are 
external to a ring but are bonded to a ring atom with a bond order greater than 
1 are  to be part of the ring system because they modify the nature of the ring. 
Atoms that are double-bonded to linker atoms are also considered to be part of 
the linker because they can modify the nature of the linker.  The basis 
scaffolds for a molecule are the set of all unique ring systems in the molecule,
where a ring system is defined as one or more rings that share an internal 
bond. 

A recursive algorithm is used to elucidate all candidate scaffold structures 
including those derived from exhaustive combinations of the basis scaffolds.   
It recursively removes each ring system from the scaffolds of each level to 
generate fragments that contain all possible ring system combinations. Finally, 
the process is completed by adding the  Bemis-Murko scaffold of the compound to 
the list of scaffold structures. 

\begin{figure}[t!]
    \centering
    \includegraphics{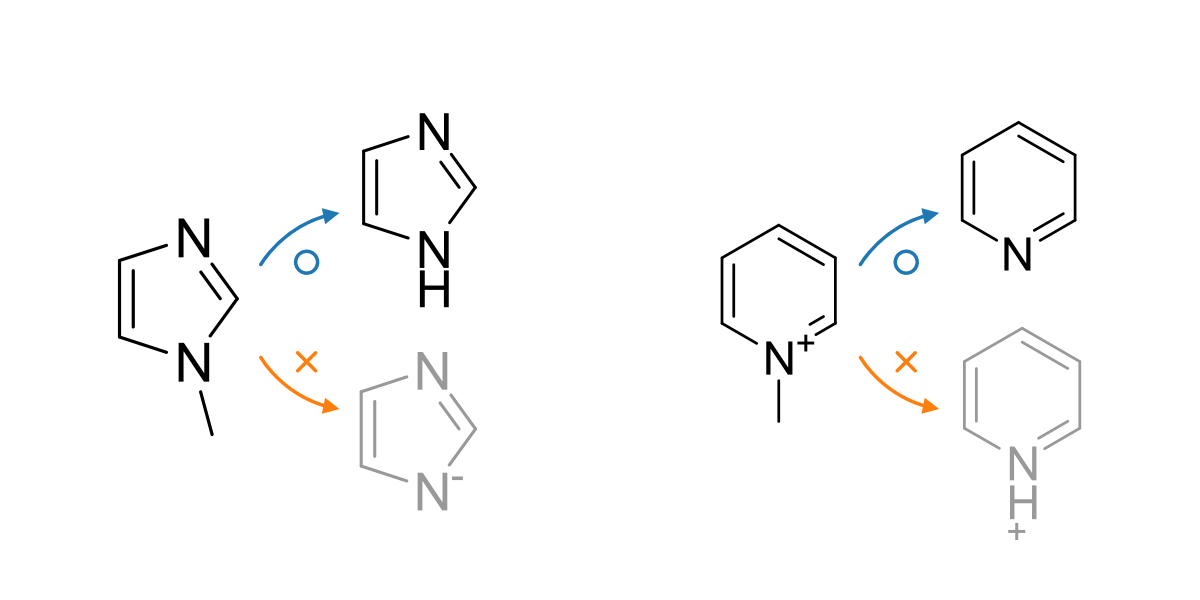}
    \caption{
        Formal charges are removed and explicit hydrogen atoms are fixed after removing chains connected with nitrogen atoms in aromatic rings.  
    }
    \label{fig:aro_n}
\end{figure}

While removing chains/linkers connected on nitrogen atoms in aromatic rings, 
things become a little complicated. We have two choices to keep a molecule in 
correct valence while fracturing such bonds (Figure \ref{fig:aro_n}). It's not 
in line with common sense about molecular scaffolds if a supplied scaffold has 
charges on some atom in it. Therefore we select the process above in Figure 
\ref{fig:aro_n} to get the scaffold without any bonds connected with nitrogen 
atoms in aromatic rings, which can be an initiation point in the scaffold-based 
generative model. The indices of processed nitrogen atoms in each molecule are 
recorded in the scaffold extraction workflow as well, so as to modify it while 
generating a molecule. 

After all the distinct basis and multi-ring system scaffolds for each molecule 
have been identified and added to the scaffold list, hierarchical structural 
relationships between the scaffolds ($s \in \mathcal{S}$) and original compound  
($m \in \mathcal{M}$) structures are established. Once all scaffolds are 
identified and their hierarchical relationships are established, the process can
rapidly traverse the network of structural connections to determine 
relationships between indices of scaffolds and its original compounds.

Finally, to facilitate the sampling of training data during model
optimization, we created indexes between skeletons
(\(c \in \mathcal{C}\)), scaffolds (\(s \in \mathcal{S}\)) and molecules
(\(m \in \mathcal{M}\)). Specifically, indices are created:

\begin{itemize}
\item
  From a given scaffold \(s\) to its corresponding skeleton \(c(s)\).
\item
  From a given scaffold \(s\) to the set of all molecules containing
  that scaffold \(\mathcal{M}(s)\)
\item
  From a given molecule \(m\) to the set of its sub-scaffolds
  \(\mathcal{S}(m)\)
\end{itemize}

%% file: sections/methods/csk.tex
\subsection{Generating scaffolds from cyclic skeletons}
\label{sec:method-csk}

As discussed in Section \ref{sec:method-overview}, to generate molecules from a
given cyclic skeleton (CSK), we need to first generate a scaffold with
atom types and bond types. To achieve this goal, we need to
model a conditional distribution \(p_{\boldsymbol\theta}(G_s|G_c)\),
where \(G_c\) is the molecular graph of CSK \(c\), and \(G_s\) is the
graph of the classical scaffold \(s\). 

\subsubsection{The generative model}

We construct the generation process as follows (see Figure \ref{fig:architecture}):

\begin{figure}[t!]
    \centering
    \includegraphics{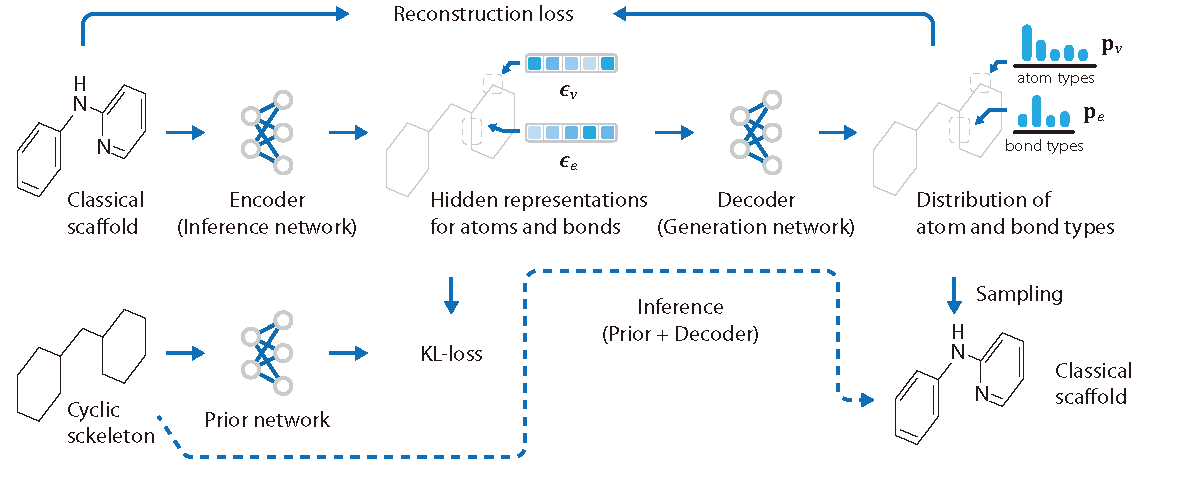}
    \caption{
        The architecture of the model to transform CSKs to classicial scaffolds.
        Under the context of autoencoder, the
        entire model contains three components: the encoder, the decoder, and
        the prior network. During training, the reconstruction loss
        \(\mathcal{L}_{\textrm{recon}}\) and the KL loss
        \(\mathcal{L}_{\textrm{KL}}\) are calculated using the encoder, decoder
        and prior for model optimization. During inference, the latent variable
        is first sampled from the prior network and then fed into the decoder
        (or the generator network) to produce the classical scaffold.
    }
    \label{fig:architecture}
\end{figure}

\begin{itemize}
\item
  We first attach to each node \(v\) and edge \(e\) in \(c\) a latent
  variable sampled from a prior distribution
  \({\boldsymbol\epsilon}_v, {\boldsymbol\epsilon}_e \sim p_{\boldsymbol\theta}({\boldsymbol\epsilon}|G_c)\).
  Here, we use a gaussian distribution with diagonal covariance
  structure to parameterize the prior:
  \(p_{\boldsymbol\theta}({\boldsymbol\epsilon}|G_c)=\mathcal{N}({\boldsymbol\epsilon}; {\boldsymbol\mu}_{\boldsymbol\theta}^\textrm{prior}(G_c), {\boldsymbol\sigma}_{\boldsymbol\theta}^\textrm{prior}(G_c)^2 \mathbf{I})\)
\item
  The latent variables
  \({\boldsymbol\epsilon}_v, {\boldsymbol\epsilon}_e \), together with
  the topological structure of the CSK, is fed into a graph
  convolutional neural network to generate a probability distribution of
  atom types for each node \(\mathbf{p}_v\) and bond types for each edge
  \(\mathbf{p}_e\). 
\item
  The types of each atom and bond in the output scaffold \(s\) are then
  sampled from the distributions \(\mathbf{p}_v\) and \(\mathbf{p}_e\).
\end{itemize}

The latent variables are necessary to capture the probabilistic
dependencies between different atoms and bonds. For convenience, we let
\({\boldsymbol\epsilon}_V = [{\boldsymbol\epsilon}_v]_{v\in V}\) ,
\({\boldsymbol\epsilon}_E = [{\boldsymbol\epsilon}_e]_{e \in E}\), and
\({\boldsymbol\epsilon}=[{\boldsymbol\epsilon}_V, {\boldsymbol\epsilon}_E]\).
Under the construction above, we can write the conditional distribution
\(p_{\boldsymbol\theta}(G_s|G_c)\) as:

\begin{equation}
  \label{eq:1}
p_{\boldsymbol\theta}(G_s|G_c) = 
\int_{{\boldsymbol\epsilon}}
{
  p_{\boldsymbol\theta}(
    G_s| G_c, {\boldsymbol\epsilon}
  )
  p_{\boldsymbol\theta}({\boldsymbol\epsilon}|G_c)
}
\end{equation}

\subsubsection{Variational auto-encoder}

A straight forward way of training such models is to maximize its
likelihood value. However, as the model contains latent variables,
calculating its likelihood value is not directly feasible. Therefore, we
use variational inference to obtain a lower-bound of the likelihood.
Specifically, we reformulate the model as a variational autoencoder
\cite{RN45} by adding an auxiliary inference network
\(q_{\boldsymbol\theta}({\boldsymbol\epsilon}|G_s, G_c)\) to the model.
Since \(G_s\) already contains the full information of \(G_c\), we shall
write the inference network as
\(q_{\boldsymbol\theta}({\boldsymbol\epsilon}|G_s)\). Similar to the
prior, we use a diagonal Gaussian distribution for
\(q_{\boldsymbol\theta}\):
\(q_{\boldsymbol\theta}({\boldsymbol\epsilon}|G_s)=\mathcal{N}({\boldsymbol\epsilon}; {\boldsymbol\mu}_{\boldsymbol\theta}^\textrm{post}(G_s), {\boldsymbol\sigma}_{\boldsymbol\theta}^\textrm{post}(G_s)^2 \mathbf{I})\).
In the context of autoencoders, the generator \(p_{\boldsymbol\theta}\)
is often referred to as "decoder", and the inference network
\(q_{\boldsymbol\theta}\) the "encoder". We then obtain the loss
function as follows:

\begin{equation}
\label{eq:2}
 \mathcal{L}({\boldsymbol\theta}) = 
-\mathbb{E}_{
	{\boldsymbol\epsilon}\sim q_{\boldsymbol\theta}
}
[
	\log{
		p_{\boldsymbol\theta}(G_s|G_c, {\boldsymbol\epsilon})
	}
] + KL(p({\boldsymbol\epsilon})||q_{\boldsymbol\theta}({\boldsymbol\epsilon}|G_s, G_c)) 
\end{equation}

The first term and second term in eq.\ref{eq:2} are usually called 
the reconstruction loss (\(\mathcal{L}_\textrm{recon}\)) and the KL
loss (\(\mathcal{L}_\textrm{KL}\)) respectively. Minimizing the loss
 eq.\ref{eq:2} is maximizing a lower bound on the likelihood
function eq.\ref{eq:1}.

\subsubsection{Graph convolutional neural network}

\begin{figure}[t!]
    \centering
    \includegraphics{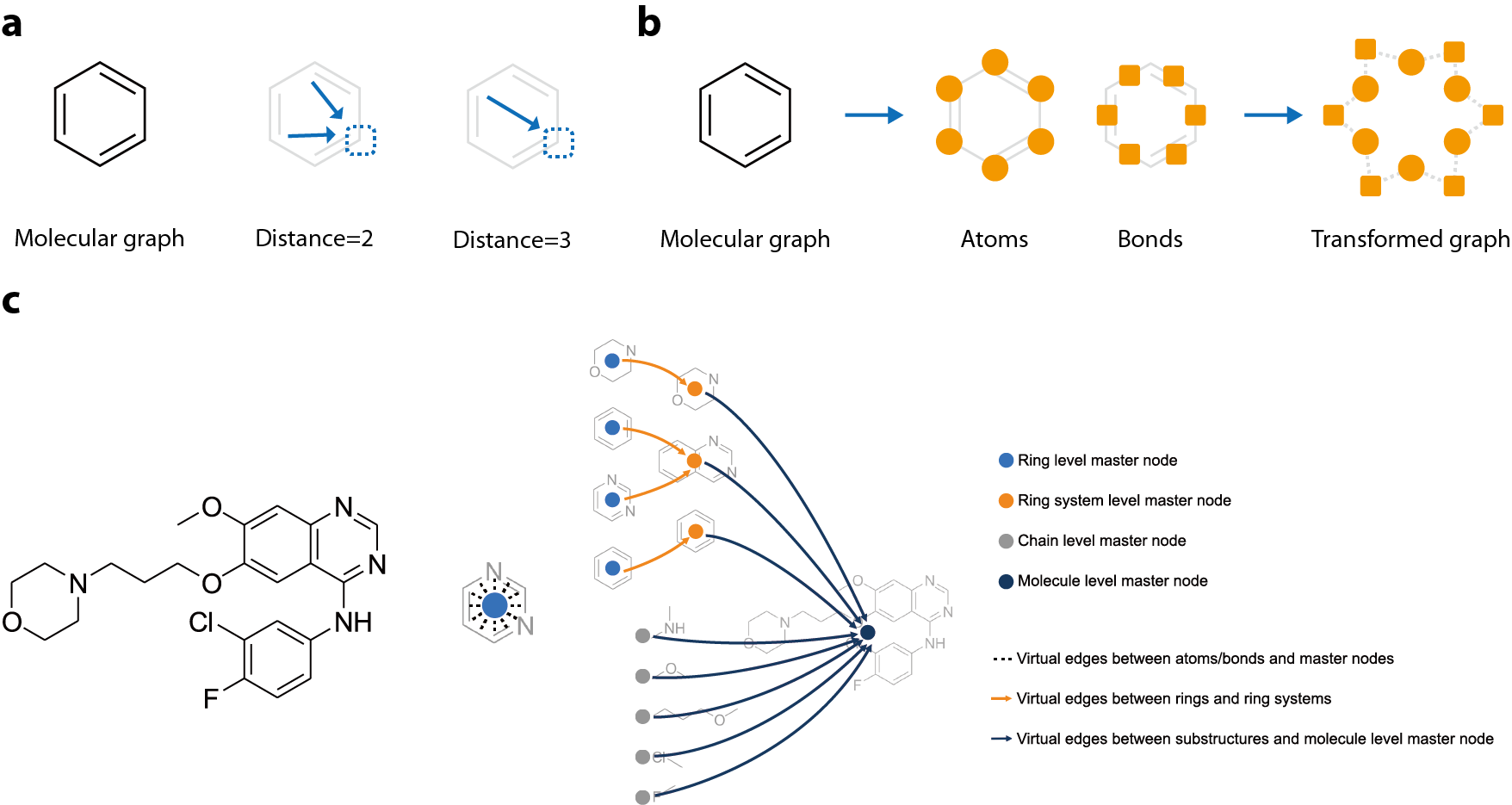}
    \caption{
        The architecture of the graph convolutional neural network (GNN).
        Before feeding into the GNN, two transformations
        are performed on the original molecular graph \(G\): \textbf{a.} Virtual
        bonds are added to capture remote connectivity relationship. \textbf{b.}
        The atoms and bonds are treated equally as "nodes" in the new graph.
        \textbf{c.} Adding a hierarchical organization to the molecular graph
    }
    \label{fig:graph-convolution}
\end{figure}

The encoder \(q_{\boldsymbol\theta}({\boldsymbol\epsilon}|G_s)\),
decoder \(p_{\boldsymbol\theta}(G_s| G_c, {\boldsymbol\epsilon})\), as
well as the prior network
\( p_{\boldsymbol\theta}({\boldsymbol\epsilon}|G_c)\) are all
parameterized using graph convolutional neural networks (GNN). The
architecture is shown in Figure \ref{fig:graph-convolution}.

Before entering the graph convolutional network, we perform the
following transformations to the input graph \(G=(V, E)\):

\begin{itemize}
\item
  Virtual bonds are added to capture remote connectivity relationship.
  Specifically, we add new bonds between atom pairs separated by two or
  three bonds, similar to our previous works
  \cite{RN236}. We also create new bond types for those
  virtual connections to separated them from real chemical
  bonds. The molecular graph with virual bonds are denoted as
  \(G^{(1)}=(V^{(1)}, E^{(1)})\).
\item
  \(G^{(1)}\) is then transformed into a new graph
  \(G^{(2)}=(V^{(2)}, E^{(2)})\) by treating atoms and bonds in
  \(G^{(1)}\) equally as nodes in \(G^{(2)}\). In other words, we have
  \(V^{(2)}=V^{(1)}\cup E^{(1)}\), and
  \(E^{(2)} = \{ \{v, e\} | v \in V^{(1)}, e \in E^{(1)} \textrm{ and } v \in e \}\).
\item
  For each ring $R$ in the graph $G$, we add a virtual node $v_R$ to the $G^{(2)}$,
  and connect it to each element inside the ring $R$ in graph $G^{(2)}$. Similar
  operations are performed for each linker in $G$. In terms of ring assemblies, 
  virtual nodes are added and are connected to virtual elements representing 
  rings. The overall effect is shown in Figure \ref{fig:graph-convolution}{\bf c.}. The
  transformed graph is denoted as $G^{(3)} = (V^{(3)}, E^{(3)})$
\end{itemize}

After those transformations, the graph $G^{(3)}$ is then sent to the graph
convolutional network as input. Each convolution operation is composed of three
stages:
\begin{itemize}
  \item
    {\bf Broadcasting:} The information of nodes are broadcasted to edges.
    Specifically, for each edge $e = \{u, v \} \in E^{(3)}$, the 
    representation of terminal nodes $\mathbf{h}_u$ and $\mathbf{h}_v$ are
    broadcasted to edge $e$ as
    \begin{equation}
      \mathbf{h}_e = \textrm{MLP}^\textrm{broadcast}([\mathbf{h}_u, \mathbf{h}_v])
    \end{equation}
    $[\cdot, \cdot]$ is the concatenation operation for vectors. 
    $\textrm{MLP}^\textrm{broadcast}$ is a linear layer with BN-ReLU-Linear 
    architecture.
  \item
    {\bf Gathering:} The information broadcasted to each edge is again 
    gathered to each node. We use to ways to gather information from edges:
    summation and maximization. The edge representation is firstly split into 
    two parts:

    \begin{equation}
    [\mathbf{h}_e^\textrm{max}, \mathbf{h}_e^\textrm{sum}] = \mathbf{h}_e
    \end{equation}

    vectors $\mathbf{h}_e^\textrm{max}$ are subjected to maximization:

    \begin{equation}
      \mathbf{h}_v^\textrm{max} = 
      \max_{e \in E^{(3)} \textrm{and} v \in e}{\mathbf{h}_e^\textrm{max}}
    \end{equation}

    vectors $\mathbf{h}_e^\textrm{sum}$ are subjected to summation:

    \begin{equation}
      \mathbf{h}_v^\textrm{sum} = 
      \sum_{e \in E^{(3)} \textrm{and} v \in e}{\mathbf{h}_e^\textrm{sum}}
    \end{equation}

    The results are concatenated:
    \begin{equation}
      \mathbf{h}_v^\textrm{gather} = [\mathbf{h}_v^\textrm{max}, \mathbf{h}_v^\textrm{sum}]
    \end{equation}
  
  \item
    {\bf Updating:} The representation of each node is updated using the 
    gathered information:
    \begin{equation}
      \mathbf{h}_v^\textrm{new} = \textrm{MLP}^\textrm{update}([\mathbf{h}_v, \mathbf{h}_v^\textrm{gather}])
    \end{equation}
    $\textrm{MLP}^\textrm{update}$ is a linear layer similar to $\textrm{MLP}^\textrm{broadcast}$
\end{itemize}

\subsubsection{Training details}

The model is validated using the holdout method. 80\% of scaffolds are used
for training (denoted as \(S^\textrm{train}\)), and the remaining 20\%
is used for testing (denoted as \(S^\textrm{test}\)). The metrics used
for test set evaluation is discussed in detail in Section \ref{sec:eval-csk}. The
mini-batches \(B=\{s_i, c_i\}_{i=1}^{|B|}\) used at each step for model
training is constructed by first sampling the scaffolds
(\(\{s_i\}_{i=1}^{|B|}\)) are then finding their corresponding cyclic
skeleton \(c_i = C(s_i)\). Naive VAE tends to suffer from the problem of
posterior collapse. To encourage the model to encode more information
into the latent variable, we experimented several attempts, including
KL-annealing (proposed in \cite{RN7}) and \(\beta\)-VAE
(proposed in \cite{RN343}).

AdaBound is used for model optimization, with the initial learning rate set
as \(1\times10^{-3}\), and the final learning rate as \(0.1\). This
means that the optimizer will gradually anneal from an Adam optimizer
with learning rate \(1\times10^{-3}\) to an SGD optimizer with learning
rate \(0.1\). The gradient is clipped to {[}-3, 3{]} to improve model
stability. The dimension of latent variables
\({\boldsymbol\epsilon}_v, {\boldsymbol\epsilon}_e\) is set to be 10.
We performed experiments with more latent variables, but found little
improvement in model performance.

%% file: sections/methods/scaffold.tex
\subsection{Scaffold-based molecule generation}
\label{sec:method-scaffold}

\begin{figure}[t!]
    \centering
    \includegraphics{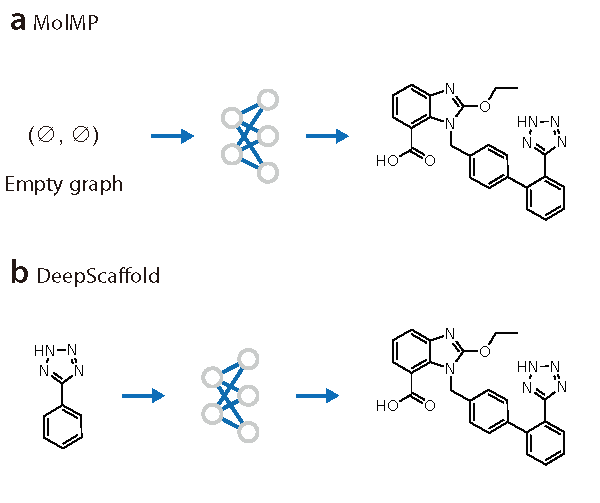}
    \caption{
        The generation process.
        \textbf{a.} Unconditional generative models (such as MolMP) starts molecule
        generation from empty graph \(G_0=(\empty, \empty)\). \textbf{b.}
        DeepScaffold starts molecule generation directly from scaffolds.
    }
    \label{fig:generation-process}
\end{figure}

Now we introduce the model used for scaffold derivatization. The central
goal of this task is to reconstruct the side-chains based on the
structural information of the scaffold. Formally speaking, we are
modeling the conditional distribution \(p_{\boldsymbol\theta}(x|s)\),
where \(s\) is the scaffold and \(x\) is the full molecule. The
generation process of side-chain follows largely from our previous work
\cite{RN236}. The major difference is that for
scaffold-based generation, the process starts from the graph of the
given scaffold, instead of an empty graph (see Figure \ref{fig:generation-process}). The model
builds the molecule in a step-by-step fashion. At each step, the model
needs to decide which of the following actions to perform: (1) append a
new atom to the graph, (2) connecting two existing nodes or (3) to
terminate the generation process. The architecture of graph
convolutional layer is similar to that used in
\cite{RN236}. In terms of the model architecture, a 20
layer densenet is used in this model, similar to Section \ref{sec:method-csk}. The
growth rate is set to 24, and the number of features in each bottleneck
layer is set to 96. We use the likelihood-based loss function for model
optimization. Importance sampling is used to improve the performance of
the model (following \cite{RN236}). The value for the
hyperparameter \(k\) (the number of samples generated from importance
sampling) and \(\alpha\) (parameter controlling the degree of uncertainty
in route sampling) is set as \(5\) and \(0.5\) respectively.

Note that our proposed model share similarity with that by Lim et. al.
Both methods share the idea of growing the molecule directly from the
scaffold. However, we improve upon the work by Lim et. al. from the following
perspectives:

\begin{itemize}
    \item
        Beside including Bemis-Murko scaffolds of
        molecules, we also included all its sub-scaffolds within our dataset
        (under the definition of HierS, see Section \ref{sec:dataset} for details). This
        treatment will have two advantages for the model. 
        \begin{itemize}
            \item
                First of all,
                expanding the scaffold set \(\mathcal{S}\) will reduce the chance of
                missing potential bioactive scaffolds, making the dataset more fitted to
                the practical need.
            \item
                Secondly, the definition of Bemis-Murko scaffold
                has excluded all ring-based structures from the side-chain, making it
                overly restricted for situations in medicinal chemistry. By including
                the sub-scaffolds, we can significantly increase the structural
                diversity of the substituents in the scaffold, thereby expanding the
                practicality of our model.
        \end{itemize}
    \item
        Compared with Lim et. al., we include a much wider range of metrics for
        model evaluation (see \ref{sec:eval}), including MMD, internal diversity, 
        docking score and so on. We also include case studies related to 
        scaffold-based drug discovery for GPCRs.
\end{itemize}

Model validation is performed in the following two ways:

\subsubsection{Holdout validation}
Holdout method is mainly used to obtain an
overall view of the model's performance. The set of all molecules is
split into two portions: the training set (80\%) and the test set
(20\%), used respectively for model optimization and evaluation. Note
that is method will result in an overlap of scaffold set between the
training and test set, therefore can not be used to examine the model's
ability to generalize to new scaffolds.

\subsubsection{Leave-one-out validation}
\label{sec:leave-one-out}
Leave-one-out validation is carried
out in the following way: Given a scaffold of interest, we gather all
molecules containing this scaffold as substructure, and exclude those
molecules from the training set. Following this method, there will be no
overlap of scaffolds and molecules between training and test set.
However, it is impractical to perform such validation to all 338,932
extracted scaffolds. Therefore, we selected three scaffolds as a case
study, as shown in Figure \ref{fig:privileged_samples}. All scaffolds selected are
privileged scaffolds for G-protein-coupled receptors (GPCRs) .

Model optimization is performed using Adam optimizer. The learning
rate is set to 0.001 in the first step, and decay at a rate of 0.99 for
every 100 steps, until the minimum learning rate \(5 \times 10^{-5}\) is
reached. Regularization techniques such as dropout or weight decay are
not used in this model. To stabilize training, the gradient is clipped to
{[}-3.0, 3.0{]} at each optimization step. Training is performed for
50,000 iterations, taking roughly 20 hours to complete.

At each step, we need to sample a mini-batch of scaffold-molecule pairs
(\(B = \{s_i, m_i\}_{i=1}^{|B|}\)) from the training set. The batch size
\(|B|\) is set to 128. Note that there are many different ways to
perform the sampling from the training set. For example, we can first
sample a set of scaffolds \(\{s_i\}_{i=1}^{|B|}\) and then assign each
scaffold \(s_i\) with a molecule drawn from \(M(s_i)\). We can also
sample molecule first and scaffold next. The impact of different
sampling algorithm is analyzed in Supporting Information. The result can be
summarized as follows: (1) If we place equal weight to each scaffold,
the molecules will be sampled disproportionately. (2) If we place equal
weight to each molecule, the scaffolds will then be sampled
disproportionately. In a word, it is quite difficult to balance both
scaffold and molecule. As a result, we construct each mini-batch
using a mixture of samples from both sampling methods. Here, we use equal weights for each sampling
method. In future works, the weight can be treated as a hyperparameter
and be subjected to optimization and tuning.

%% file: sections/methods/pharmacophore.tex
\subsection{Using scaffold and pharmacophore as queries for generation}
\label{sec:method-pharmacophore}

For medicinal chemists, it is would often be useful to have the ability to 
specify the property of substituents at specific locations during 
scaffold-based molecule design. This may happen when we have clear SAR 
information concerning the given scaffold. In this work, we also created a 
tool to use both scaffold and pharmacophore queries during molecule generation. 
The implementation of this functionality is rather simple: after the molecules 
are generated from the scaffold using the method discussed in Section \ref{sec:method-scaffold}, 
we filter those results against the user's queries. We currently allow users 
to specify the size of the side-chain (measured using the number of heavy atoms
), whether to have hydrogen bond donors (HBD) and whether to have hydrogen bond 
acceptors (HBA). We plan to increase the spectrum of supported queries in the 
future.

%% file: sections/methods/evaluation.tex
\subsection{Evaluation}
\label{sec:eval}
\subsubsection{Generating scaffolds from cyclic skeletons}
\label{sec:eval-csk}

We use the following metrics for model evaluation:

\textbf{Reconstruction loss, KL loss and total loss}
VAE learn the generative model by trying the achieve the best
reconstruction using the least information. The reconstruction loss
\(\mathcal{L}_\textrm{recon}\) measures the accuracy of reconstruction
for the autoencoder, while the KL loss \(\mathcal{L}_\textrm{KL}\)
measures the amount of information stored inside the latent variable.
The total loss \(\mathcal{L}_\textrm{total}\), which is also used during
the model training, is the sum of reconstruction loss and KL loss.

\textbf{Chemical validity}
Given a cyclic skeleton \(c\), we measure the chemical validity of
output scaffolds generated from skeleton \(c\). The validity of a given
molecule is evaluated using the \texttt{Chem.Sanitize} API provided by
RDKit. The validity is measured by the percent valid output calculated
as follows:

\begin{equation}
  P_c^\textrm{valid} = N_c^\textrm{valid}/N_c^\textrm{total}
\end{equation}

where \(N_c^\textrm{valid}\) is the number of valid outputs and
\(N_c^\textrm{total}\) is the number of total outputs. Those values are
averaged among all cyclic skeletons, and the final value
\(\bar{P}_\textrm{valid}\) is reported.

\subsubsection{Evaluating scaffold-diversification model}

In order to evaluate the performance of the scaffold-based molecular
diversification model, we gather all scaffolds occurred inside the test
set, and remove those with less than 10 corresponding molecules. After
obtaining the set of scaffolds to be evaluated (denoted as
\(S_\textrm{test}\)), 10,000 sample molecules are generated for each
scaffold, and the evaluation of those samples is carried out using the
following metrics.

\textbf{Chemical validity and uniqueness}
We first measure the chemical validity and uniqueness of the generated
molecules. The validity of a given molecule is evaluated using the
\texttt{Chem.Sanitize} API provided by RDKit. For a given scaffold
\(s\), we obtaining the number of valid outputs (denoted as
\(N_s^\textrm{valid}\)), and divide it with the total number of
generated molecules (denoted as \(N_s^\textrm{total}\)). The percentage
of valid outputs are calculated using eq.\ref{eq:5}.

To calculate the uniqueness, we remove all duplicated molecules from the
valid outputs, and calculate the number of remaining molecules (denoted
as \(N_s^\textrm{uniq}\)). The uniqueness is calculated using
eq.\ref{eq:6}.

\begin{equation}
   \label{eq:5}
   P_s^\textrm{valid} = N_s^\textrm{valid}/N_s^\textrm{total}
\end{equation}
\begin{equation}
  \label{eq:6}
   P_s^\textrm{uniq} = N_s^\textrm{uniq}/N_s^\textrm{valid}
\end{equation}

\(P_s^\textrm{valid}\) and \(P_s^\textrm{uniq}\) scores are calculated
for each scaffold inside the test dataset, and the average values of the
two metrics (\(\bar{P}_\textrm{valid}\) and \(\bar{P}_\textrm{uniq}\))
are reported. The distribution of the two metrics among all test set
scaffolds is also reported. Also, we will discuss how scaffold
sizes would affect those two performance metrics.

\textbf{Distribution of molecular properties}
\label{sec:eval-prop}
Next, we investigate the model's ability to correctly model the
distribution of several important molecular properties. Here we consider
the following three properties: molecular weight (denoted as \(MW\)),
log partition coefficient (denoted as \(\textrm{log}P\)) and the
drug-likeness score (measured by QED \cite{RN26}). All
properties above can be calculated using RDKit.

For each scaffold \(s\) in the test set, we compare the mean and
standard deviation of each property (\(MW\), \(\textrm{log}P\) and
\(QED\)) between the molecules generated and molecules in the test set.
Previous studies have used metrics such as \(D_{KL}\) and \(D_{JD}\) to
quantify the difference. The approach adopted here is simpler, but offers
higher interpretability compared with \(D_{KL}\) and \(D_{JD}\). For a
quantitative evaluation, we instead use maximum mean discrepancy (MMD),
as discussed in Section \ref{sec:mmd}.

\textbf{Diversity}
\label{sec:eval-diversity}
Output diversity is another important metric for evaluating molecular
generative models. Although, as discussed in Section \ref{sec:eval-prop}, the standard
deviation of molecular properties could, in some extent, reflect the
structural diversity of the generated outputs, a more rigorous
definition of structural diversity is needed for quantitative
evaluation. Here, we adopt the formulation of internal diversity
proposed by Benhenda et. al\cite{RN344}. Supposed the
molecules are sampled from a molecule distribution \(x \sim p(x)\), the
structural diversity of this distribution can be defined as:

\begin{equation}
  I = 1 - \mathop{\mathbb{E}}_{x, y \sim p}[k(x, y)]
\end{equation}

Where \(k(x, y)\) is the similarity of the two molecules \(x\) and
\(y\). Here, we use the tanimoto score of the Morgan fingerprints
between the two molecules as the similarity measurement. Intuitively
speaking, the structural diversity of \(p\) is defined as one minus the
average similarity between two randomly sampled molecules from \(p\). We
can estimate the value of \(I\) from samples \(\{x_i\}_{i=1}^N\) drawn
from \(p\):

\begin{equation}
  \hat{I}_U = 1 - \frac{\sum_{i=1}^N\sum_{j=1, j\ne i}^N k(x_i, x_j)}{N(N-1)}  
\end{equation}

It can be proved that \(\hat{I}_U\) is an unbiased estimator of \(I\).
Note that this estimator is slightly different from that originally
proposed by Benhenda et. al. \cite{RN344}:

\begin{equation}
  \hat{I}_V = 1 - \frac{\sum_{i,j=1}^Nk(x_i, x_j)}{N^2}
\end{equation}

It can be shown that \(\hat{I}_V\) is equals to
\(\frac{N - 1}{N}\hat{I}_U\), and therefor a biased V-staticstics for
\(I\). Also, it can be shown that the bias of \(\hat{I}_V\) depends on
the sample size \(N\), and will be larger in smaller datasets. This
could be problematic for two ways: First, many scaffolds in the test set
have only a small number of corresponding molecules, which may increase
the bias. Second, the sample size (\(N\)) of the test set is usually
smaller than that of the generated samples, which will bring additional
bias to the result. Therefore, the U-statistics \(\hat{I}_U\) is instead
used in this paper for the calculation of internal diversity.

\textbf{Maximum mean discrepancy (MMD)}
\label{sec:mmd}
The central task of scaffold-based diversification is to model the
conditional distribution \(p(x|s)\) of molecular structure \(x\) given
the scaffold \(s\). Therefore, the most natural way of evaluating such
models would be measuring the dissimilarity between the ground truth
distribution \(p_d(x|s)\) and the modeled distribution \(p_g(x|s)\).
Maximum mean discrepancy (MMD) is one of such metrics that have been
widely applied in generative modeling (both for model training
\cite{RN345} and evaluation
\cite{RN197}). Different from other metrics, such as
Jensen--Shannon divergence (JSD), Wasserstein distance (WD)
\cite{RN180} or Frechet ChemNet distance (FCD)
\cite{RN245}, MMD does not require additional
discriminator or preditor to be evaluated, which makes it much
easier to calculate.

Given the samples from two probability distribution
\(\{x_i\}_{i=1}^N \sim p_d\), \(\{y_i\}_{i=1}^M \sim p_g\), the MMD
between the two distribution can be estimated as:

\begin{equation}
  \hat{D} = \frac{1}{N(N-1)}\sum_{i=1}^N\sum_{j=1, j \ne i}^Nk(x_i, x_j) + \frac{1}{M(M-1)}\sum_{i=1}^M\sum_{j=1, j \ne i}^Mk(y_i, y_j) - \frac{2}{MN}\sum_{i=1}^N\sum_{j=1}^Mk(x_i, y_j)
\end{equation}

Where \(k(\cdot, \cdot)\) is the kernel function. Similar to
Section \ref{sec:eval-diversity}, we choose the tanimoto similarity between molecules as
\(k\). It can be proved that tanimoto similarity is in fact a legitimate
kernel function \cite{RN346}.

One of the most important applications the model is the design of
molecule-based on privileged scaffolds. Ideally, given a known bioactive
scaffold, the generated molecule should also be bioactive. To
test whether the model possess such desirable property, we examine:

\begin{itemize}
\item
  The ability for the generative model to reproduce known active
  molecules
\item
  The (predicted) bioactivity of the output molecules for the target.
\end{itemize}

\textbf{The rate of reproduced active compounds}
First of all, the rate reproduced active molecules (denoted as \(R\))
for GPCRs is reported. Specifically, we collect all molecules that are
active against at least one target in the GPCR family (denoted as the
set \(T_\textrm{GPCR}\)), and examine how many of them appear in the
set of molecules sampled from the model(denoted as \(M_s\)):

\begin{equation}
  R_\textrm{GPCR} = \frac{|M_s \cap T_\textrm{GPCR}|}{|T_\textrm{GPCR}|}
\end{equation}

We also investigate the ability for the model to reproduce known drugs,
by collecting all drugs (denoted as \(T_\textrm{drug}\)) from the test
set, and calculate the ratio \(R_\textrm{drug}\):

\begin{equation}
  R_\textrm{drug} = \frac{|M_s \cap T_\textrm{drug}|}{|T_\textrm{drug}|}
\end{equation}

\textbf{Docking}
Next, we report the predicted activity of output molecule for GPCRs. It
is impractical to evaluate the output molecule against all targets in
the GPCR family. Instead, we choose Dopamine Receptor D2 (DRD2) as a
representative, which is a well-studied target for antipsychotic drugs
\cite{RN347}. The predicted activity for DRD2 is
calculated using docking. The bioactivity model for DRD2 is built
following Supporting Information:

For each bioactive scaffold selected for GPCRs, a comparison of docking
scores was carried out between four sets of compounds: (1) samples
generated by the model, (2) molecules inside the test set, (3) known
actives for DRD2 and (4) randomly sampled molecules from ChEMBL. A well-performing 
generative model should result in a similar distribution of
docking scores between (1) and (2). Also, the average score for (1)
should be close to (3), and better than (4).

\textbf{Structural distribution of side-chains}
Given the set of generated molecules given a scaffold, users can further
filter the result by adding requirement on the properties of side-chains
attached to the scaffold. Since different users may require different
types of substitution in different locations, the model needs to generate
a diverse set of side-chain configurations in order to handle the user's
requirements efficiently. Therefore, for each scaffold in the
leave-one-out evaluation (see Section \ref{sec:leave-one-out}), we analyze the location
and size of each side-chain in the generated molecules in order to
evaluate the diversity of side-chains. We also compare the result with
test set molecules to determine the quality of the generated samples.

%% file: sections/results.tex
\section{Results}
\input{sections/results/overview.tex}
\input{sections/results/csk.tex}
\input{sections/results/scaffold.tex}

%% file: sections/results/overview.tex
\subsection{Generated samples}

\begin{figure}[b!]
\centering
\includegraphics{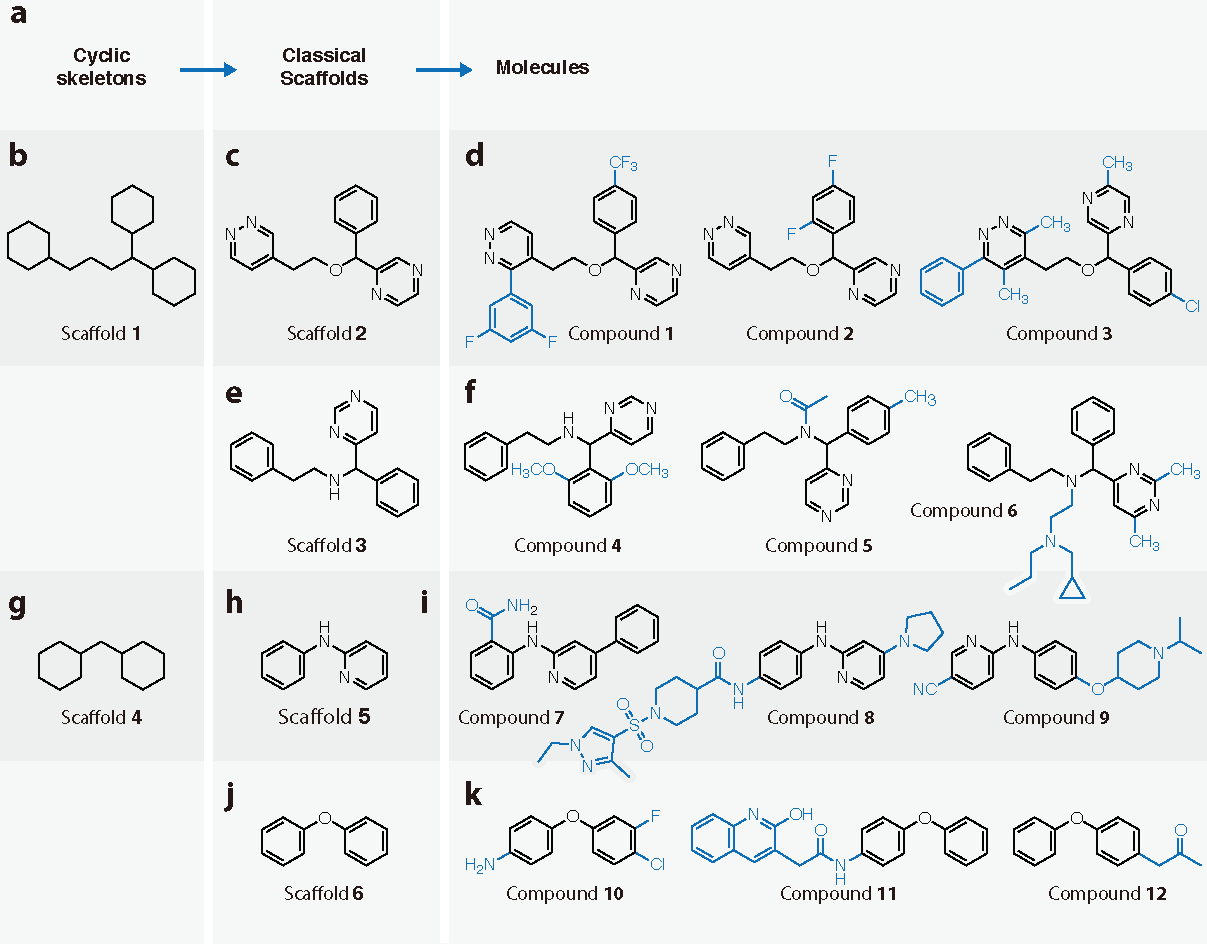}
\caption{Generated samples from DeepScaffold.}
\label{fig:samples}
\end{figure}

Before diving into model's performance, we present a brief view of
DeepScaffold by showcasing several generated molecules from cyclic
skeletons as well as classical scaffolds using the model. The result is
demonstrated in Figure \ref{fig:samples}. Two cyclic skeletons, scaffold
\textbf{4} (Figure \ref{fig:samples} \textbf{g}), with a relatively
simple structure, as well as the more complexed scaffold \textbf{1}
(Figure \ref{fig:samples} \textbf{b}), are selected from the hold-out test
set. For each scaffold, we generate two classical scaffolds (scaffold
\textbf{2} (Figure \ref{fig:samples} \textbf{c}) and scaffold
\textbf{3} (Figure \ref{fig:samples} \textbf{e}) for scaffold
\textbf{1}, as well as scaffold \textbf{5} (Figure \ref{fig:samples}
\textbf{h}) and scaffold \textbf{6} (Figure \ref{fig:samples} \textbf{j})
for scaffold \textbf{4} and structural derivification is performed
on those structures (Figure \ref{fig:samples} \textbf{d,f,i,k}).

%% file: sections/results/csk.tex
\subsection{Generating scaffolds from cyclic skeletons}

The performance of the proposed model is evaluated following
Section \ref{sec:eval-csk}, with results shown in Table \ref{tab:csk}.

First of all, it is shown that approaches such as KL-annealing and
\(\beta\)-VAE all showed positive results in solving posterior collapse.
Both VAE + annealing and \(\beta\)-VAE are capable of achieving higher
\(\mathcal{L}_\textrm{KL}\). Notably, by using annealing, the model can achieve 
a slightly better $\mathcal{L}_\textrm{total}$ with a much higher 
$\mathcal{L}_\textrm{KL}$. In terms of the output validity, \(\beta\)-VAE (with 
$\beta$ value set to 0.5) can achieve the best performance of 82.5\%. 
Interestingly, the output validity is not always consistent with the performance
measured using $\mathcal{L}_\textrm{total}$. In the future, we plan to utilize more
recent techniques in VAE training in order to improve the result, such as
using more complexed prior \cite{RN38} posterior \cite{RN22}.

\begin{longtable}[t!]{@{}llllll@{}}
    \caption{Generating scaffolds from cyclic skeletons}\label{tab:csk}\\
    \toprule
    Model & \%valid & \%recon & \(\mathcal{L}_\textrm{KL}\) &
    \(\mathcal{L}_\textrm{recon}\) &
    \(\mathcal{L}_\textrm{total}\)\tabularnewline
    \midrule
    \endhead
    \(\beta\)-VAE (\(\beta=0.5\)) & 82.50\% & 61.00\% & 9.82 & 1.11 &
    10.93\tabularnewline
    \(\beta\)-VAE (\(\beta=0.\)1) & 80.70\% & 91.30\% & 14.15 & 0.23 &
    14.38\tabularnewline
    VAE + KL annealing & 76.80\% & 10.00\% & 5.02 & 4.38 &
    9.4\tabularnewline
    VAE & 68.50\% & 6.32\% & 3.37 & 6.06 & 9.43\tabularnewline
    \bottomrule
\end{longtable}

%% file: sections/results/scaffold.tex
\subsection{Scaffold derivation}
\input{sections/results/scaffold/validity.tex}
\input{sections/results/scaffold/props.tex}
\input{sections/results/scaffold/diversity-mmd.tex}
\input{sections/results/scaffold/analysis.tex}

\input{sections/results/scaffold/case-study.tex}
\input{sections/results/scaffold/side-chain.tex}
\input{sections/results/scaffold/docking.tex}

%% file: sections/results/scaffold/validity.tex
\subsubsection{Validity and uniqueness}

\begin{figure}[t!]
    \centering
    \includegraphics{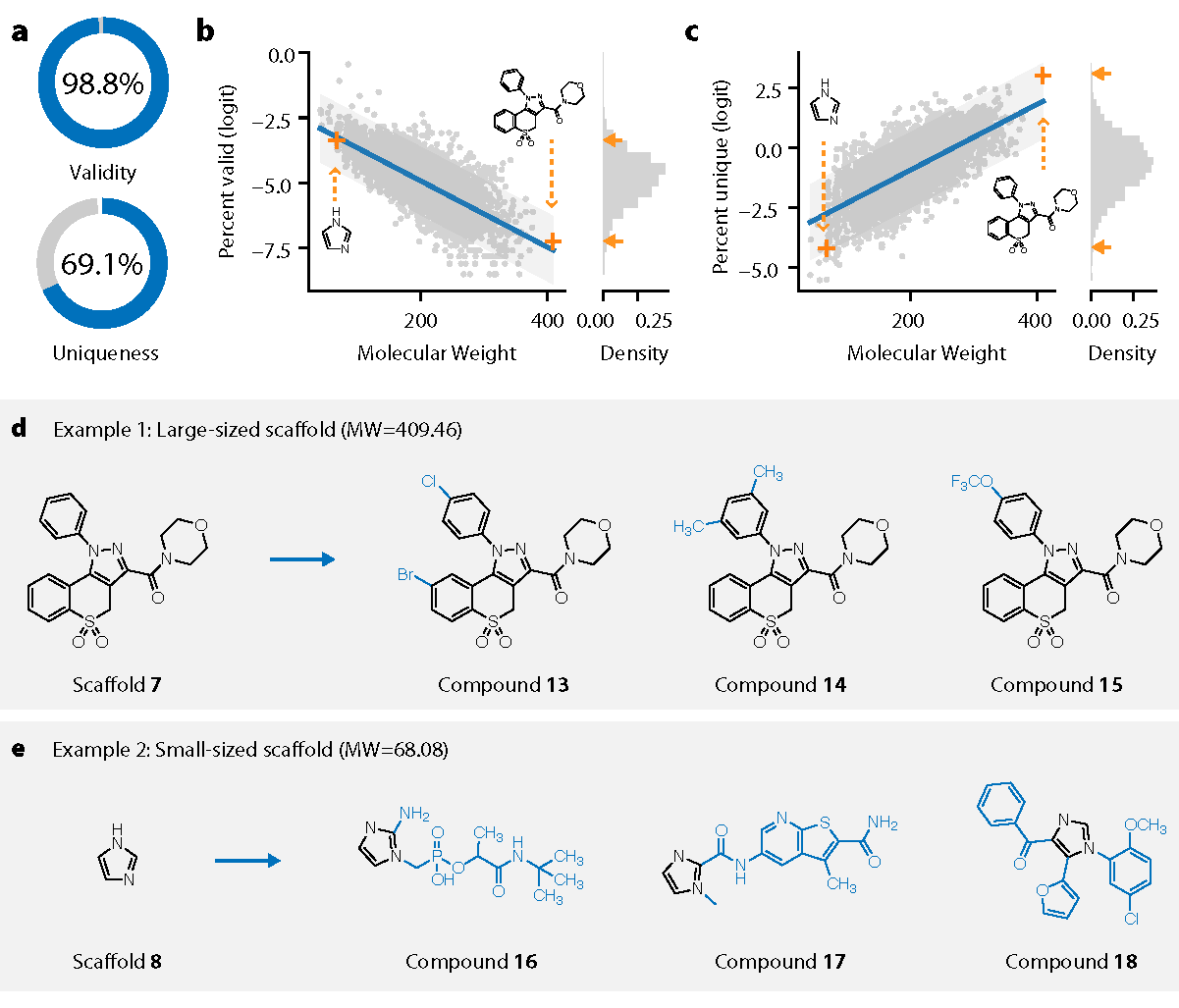}
    \caption{
        Sample validity and uniqueness.
        \textbf{a.} The average validity (98.8\%) and uniqueness (69.1\%) among
        all scaffolds. \textbf{b.} The relationship between the molecular weight of
        each scaffold and the (logit) precent-valid output. \textbf{c.} The
        relationship between molecular weight for each scaffold and (logit)
        percent unique output. \textbf{d-e.} Several generated samples using
        scaffolds with \textbf{d.} high molecular weight and \textbf{e.} low
        molecular weight.
    }
    \label{fig:valid_uniq.pdf}
\end{figure}

The performance of scaffold-based molecular diversification in terms of
validity and uniqueness is summarized in Figure \ref{fig:valid_uniq.pdf}. Among all
scaffolds, the average percentage of valid output
(\(\bar{P}_\textrm{valid}\)) is 98.9\%, and the average percentage of
unique output is 69.1\% (\(\bar{P}_\textrm{uniq}\)), as shown in
Figure \ref{fig:valid_uniq.pdf}. It is immediately noticed that the result in
\(\bar{P}_\textrm{valid}\) and \(\bar{P}_\textrm{uniq}\) is
significantly different from the results produced by the unconditional
version of the model with similar architecture \cite{RN236}.
Specifically, the scaffold-based model can achieve a higher precent-valid
value, while having a lower percent unique value. A similar result was
reported by Lim et. al. \cite{RN339} using a different dataset
including BM scaffolds.

To better understand the phenomenon, we further investigate the
relationship between the scaffold size (measured by the molecular weight
of the scaffold \(MW_s\)) and the performance of the two metrics. The
result is shown in Figure \ref{fig:valid_uniq.pdf} and Figure \ref{fig:valid_uniq.pdf}. It can be shown that
\(MW_s\) is positively correlated with \(P_s^\textrm{valid}\)
(correlation coefficient \(r=\)0.73, significance level \(p < 0.001\) ),
while negatively correlates with \(P_s^\textrm{uniq}\) (correlation
coefficient \(r=\)-0.66, significance level \(p < 0.001\)). This means
that larger scaffolds tend to have higher output validity, but lower
uniqueness. The result is in fact inline with the mode of thinking in
medicinal chemistry: for smaller scaffold, there is usually a plenty of
room for structural diversification, while for larger scaffolds, in
order to achieve higher drug-likeness and synthetic accessibility, the
size and diversity of structural modification are much more restricted,
resulting in a higher chance of duplicates. Also, smaller side-chains means
fewer generation steps, which will, therefore, reduce the chance of
validity violation during generation. Two representative scaffolds are
chosen from the test scaffold set and are shown in Figure \ref{fig:valid_uniq.pdf} and
Figure \ref{fig:valid_uniq.pdf}. Consistent with the analysis above, the side-chains
attached to the smaller scaffold is larger and more diverse, and those
attached to the larger scaffold have a much simpler structure.

The discussion above also provides some important insight into how
future benchmarks should be constructed for this type of model. First
of all, since the evaluation metrics such as \(P_s^\textrm{valid}\) and
\(P_s^\textrm{uniq}\) depends on the scaffold \(s\) chosen for
evaluation, it is important to use a common scaffold set as the test set
during model comparisons. Secondly, it will be better to perform the
evaluation in a per-scaffold fashion. Specifically speaking, instead of
comparing the average validity and uniqueness between models, it will be
better to perform paired tests of those metrics.

%% file: sections/results/scaffold/props.tex
\subsubsection{Molecular properties}
\label{sec:result-prop}

\begin{figure}[t!]
    \centering
    \includegraphics{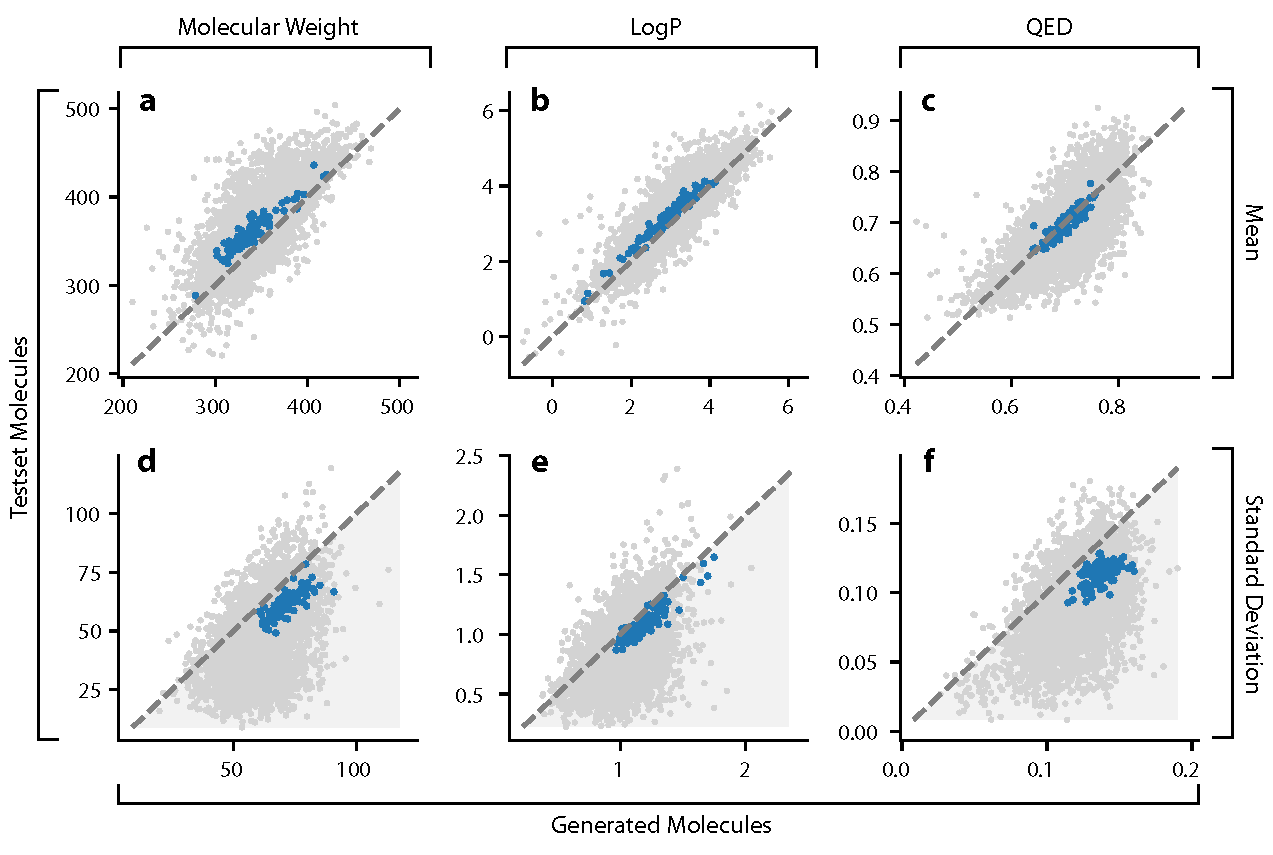}
    \caption{
        Comparing the distribution of several
        molecular properties (MW, LogP and QED) between generated and test set
        molecules.
        For each subplot, the horizontal coordinate indicates the
        statistics (\textbf{a-c} for mean and \textbf{d-f} for standard
        deviation) of a molecular property (\textbf{a} and \textbf{d} for MW,
        \textbf{b} and \textbf{e} for LogP, \textbf{c} and \textbf{f} for QED)
        of generated molecules, while its y coordinate indicates that statistics
        for the test set molecules. Scaffolds with more than 400 corresponding
        molecules (that is, \(|\mathcal{M}(s)|>400\)) are highlighted in blue.
    }
    \label{fig:props}
\end{figure}

Next, the distributions of molecular properties for generated molecules
are analyzed and compared with that in the test set. The results are
summarized in Figure \ref{fig:props}. Each point in the scatter point corresponds
to a scaffold in the test set. Its horizontal coordinate indicates the
statistics (mean or standard deviation) of a molecular property (MW,
LogP or QED) of generated molecules, while its y coordinate indicates
that statistics for the test set molecules. Scaffolds having more than
400 corresponding molecules in the test set are lighted in blue.

We first examine the mean statistics of molecular properties. Form
Figure \ref{fig:props}, it can be observed that, for all three properties, the
points in the scatter plot lies relatively close to the diagonal line,
demonstrating that the model is performing well in modeling the means of
those properties. It is also noted that scaffolds with more molecules
(highlighted in blue) lies much closer to the diagonal line compared
with other scaffolds. This may indicate better performance in
scaffolds with more data points.

However, in terms of standard deviation, there is a large discrepancy
between the generated and test samples. It is observed that a large
number of data points lies below the diagonal line, which indicates that
the generated molecule usually have a more dispersed distribution of
molecular properties. This is ,in fact, a result of higher structural
diversity of output molecules, as is shown in Section \ref{sec:result-diversity}. Again,
highlighted points are observed to lie much closer to the diagnal line,
showing that the model performs much better on those scaffolds compared
with the rest.

%% file: sections/results/scaffold/diversity-mmd.tex
\subsubsection{Molecule diversity}
\label{sec:result-diversity}

We subsequently analyze the internal diversity of the sampled and test
set molecules, with results summarized in Figure \ref{fig:diversity}. Similar to the
result showed in Figure \ref{fig:diversity}, we have observed that most data points
lie below the diagonal line, indicating a higher structural diversity of
sampled molecules compared with test set molecules. Also, a similar
result is observed that scaffolds with more data in the training set
perform much better compared with other scaffolds (see the data points
highlighted in blue).

\subsubsection{MMD}

\begin{figure}[b!]
    \centering
    \includegraphics{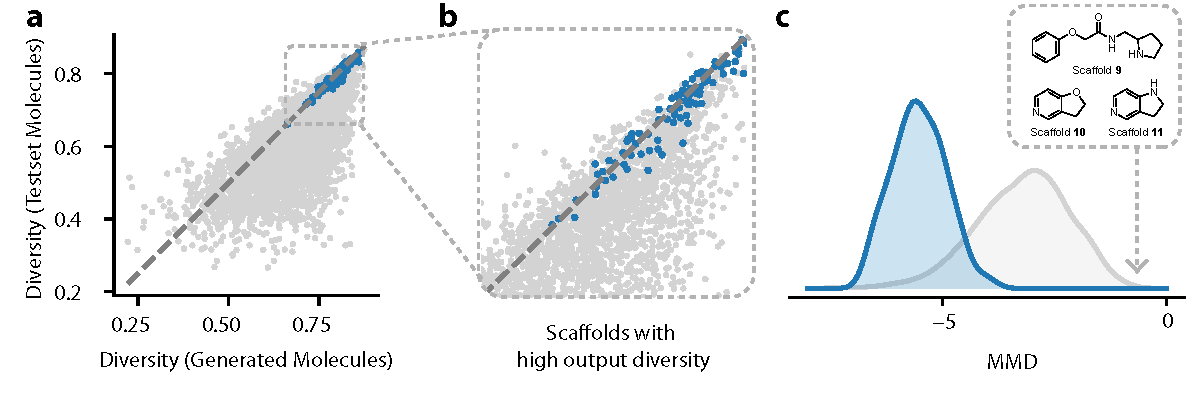}
    \caption{
        Internal diversity and MMD.
        \textbf{a.}
        Comparing the internal diversity between generated molecules and testset
        molecules. Scaffolds with more than 400 corresponding molecules (that
        is, \(|\mathcal{M}(s)|>400\)) are highlighted in blue. \textbf{b.}
        Scaffolds with high output diversity. \textbf{c.} The distribution of
        MMD among all scaffolds in the test set. Scaffolds with more than 400
        corresponding molecules are highlighted in blue.
    }
    \label{fig:diversity}
\end{figure}

The result for MMD is shown in Figure \ref{fig:diversity}. The grey curve shows the
distribution of MMD among all test set scaffolds, while the blue curve
shows the distribution of MMD among scaffolds with more than 400 related
molecules. It can be immediately observed from the figure that the model
performs significantly better for scaffolds included in the blue line.

%% file: sections/results/scaffold/analysis.tex
\subsubsection{Analyzing bad-cases}\label{sec:bad-case}

In summary, the evaluation results reported in the above sections 
lead to the following conclusions regarding the model'a performance:

\begin{itemize}
    \item
        With a given scaffold, molecules generated by the model tend to be
        more structurally diverse compared with the molecules containing that
        scaffold in the test set
    \item
    The distribution of generated molecules matches better with that of
    the test set molecules for scaffolds that occurs more frequently.
\end{itemize}

\begin{figure}
    \centering
    \includegraphics[scale=0.9]{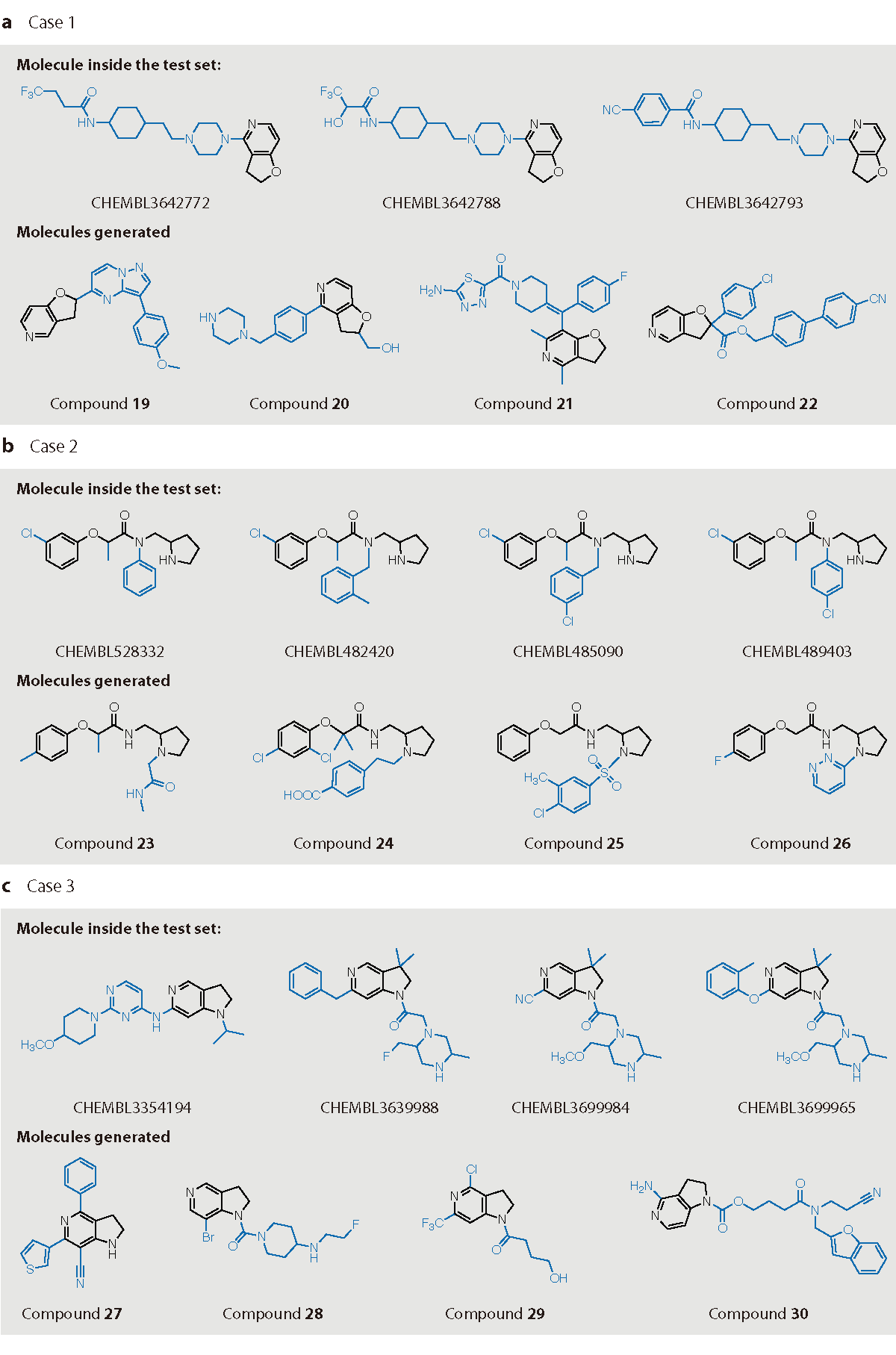}
    \caption{
        Demonstration of several bad-cases drawn from the test set scaffolds.
    }
    \label{fig:bad-case}
\end{figure}

In order to get a better understand those phenomenons, we performed a
bad-case analysis considering scaffolds with lowest MMD values, as
demonstrated in Figure \ref{fig:bad-case}. Generated molecules for those scaffolds
are shown in Figure \ref{fig:bad-case}, while molecules containing those scaffolds in
the test set are shown in Figure \ref{fig:bad-case}.

As we examine the results in Figure \ref{fig:bad-case}, we can immediately notice the
low structural diversity for the test set molecules. Interestingly, after
inspecting the origin of those molecules in the ChEMBL dataset, we found
that those molecules are mostly collected from a single publication or a
single assay: All molecules containing scaffold
\textbf{9} is originated from the assay CHEMBL3707834,
and all molecules containing scaffold \textbf{10} is
originated from the assay CHEMBL3705917. 89\% of molecules containing
scaffold \textbf{11} is collected from the publication
CHEMBL1144160. In other words, for each scaffold shown in Figure \ref{fig:bad-case},
its corresponding molecules are largely designed by the same group of
researchers under a same target.

The observations above can offer an explanation to the seemly bad
performance of the model. It can be concluded that the test set for scaffold
\textbf{9-11} suffers from the
non-random-sampling bias: the molecules is sampled from a much more
restricted chemical space that reflects the preference of specific 
research or publication. Therefore, the performance of the model will be
underestimated, resulting in high MMD values, as well as large
discrepancy between structural and property distributions. On the
contrary, a higher diversity among generated molecule is a sign
that the model did not memorize the side-chain distribution for each
scaffold, indicating high generalizability.

The foregoing discussion proposes a challenge for model evaluation in
scaffold-diversification tasks. In order to obtain a good assessment for
the model's performance, it is necessary to use a test set that is
unbiased in molecule distribution. Future research may explore the
potential of developing an unbiased benchmark set for
scaffold-diversification models. Another solution would be introducing
sample weight during evaluation to mitigate the non-random-sampling
bias, which will be left for future researches.

%% file: sections/results/scaffold/case-study.tex
\subsubsection{Case study: privileged scaffolds for GPCRs}

\begin{figure}
    \centering
    \includegraphics{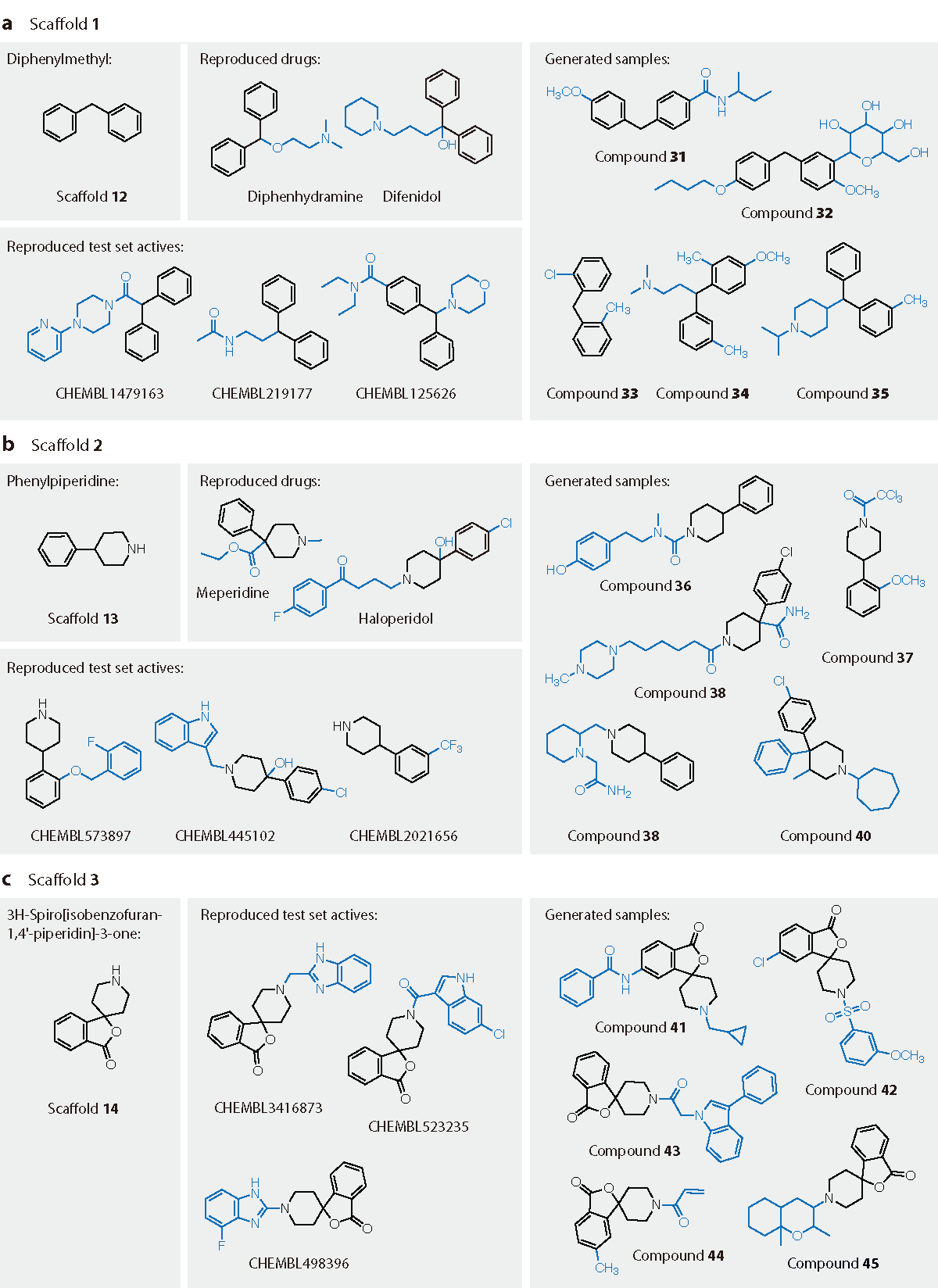}
    \caption{
        Generated samples and reproduced active molecules.
        This figure shows the generated molecules by the model given
        each scaffold (\textbf{a.} for scaffold \textbf{12},
        \textbf{b.} for scaffold \textbf{13}, \textbf{c.} for
        scaffold \textbf{14}) as well as the reproduced drugs and
        active molecules.
    }
    \label{fig:privileged_samples}
\end{figure}

As described in Section \ref{sec:leave-one-out}, three privileged scaffolds for GPCRs
(shown in Figure \ref{fig:privileged_samples}) are selected for the case study. We performed
leave-one-out evaluation for each scaffold, with the results summarized
in Table \ref{tab:scaffold} and Figure \ref{fig:privileged_samples}.

The model performs well on the percentage of valid outputs, with values
closing to 100\%. The uniqueness for scaffold \textbf{12}
and \textbf{13} is 86.6\% and 87.7\%
respectively, while for \textbf{14}, the value is of
45.4\%. This smaller value might result from the
higher structural complexity of scaffold \textbf{14},
which may restrict the chemical space of the side chains. Similar to the
results demonstrated in Section \ref{sec:result-prop} and Section \ref{sec:result-diversity} generated
molecules have a wider distribution of MW, LogP and QED, and is more
structurally diverse compared to real samples. In terms of MMD, we
found that scaffold \textbf{14} have a significantly
lower MMD value compared with scaffold \textbf{12} and
\textbf{13}. The high MMD value can largely be
contributed to the non-random-sampling bias in the test set. In fact,
most scaffold containing scaffold \textbf{14} is
collected from two publications (ChEMBL id: CHEMBL1008329,
CHEMBL1001157), with both publications focusing on the same target
called Neuropeptide Y receptor type 5 (NPY Y5).

Next, we investigate the ability for the model to reproduce know active
molecules against GPCRs (see results in Table \ref{tab:scaffold}). Within 10,000
samples, the model can reproduce 5.17\% active
molecules and 12.8\% known drugs containing scaffold
\textbf{12}, 8.44\% active molecules
and 25\% known drugs containing scaffold
\textbf{13}, as well as 15.9\%active
molecules containing scaffold \textbf{14}. Several reproduced drugs are shown in
Figure \ref{fig:privileged_samples}. Since the model has no access to those 
molecules during the training phase, the results above demonstrated good potential for
drug discovery based on privileged scaffolds.

\begin{longtable}[b!]{@{}llll@{}}
    \caption{Case study: privileged scaffolds for GPCRs}\label{tab:scaffold}\\
    \toprule
    & Scaffold \textbf{12} & Scaffold \textbf{13} & Scaffold \textbf{14} \tabularnewline
    \midrule
    \endhead
    Diversity & 0.166 & 0.187 &
    0.485\tabularnewline
    Molecular Weight & 363.6 (58.1) &
    384.9 (64.6) & 385.3
    (38.0)\tabularnewline
    LogP & 4.04 (1.06) & 4.01
    (1.02) & 3.72
    (0.698)\tabularnewline
    QED & 0.675 (0.114) & 0.708
    (0.121) & 0.660
    (0.0492)\tabularnewline
    MMD & 0.0185 & 0.00875 &
    0.154\tabularnewline
    R(actives) & 0.0517 & 0.0844 &

    0.159\tabularnewline
    R(drugs) & 0.128 & 0.25 & -\tabularnewline
    \(P_s^\textrm{valid}\) & 0.986 & 0.974 & 0.989\tabularnewline
    \(P_s^\textrm{uniq}\) & 0.866 & 0.877 & 0.454\tabularnewline
    \bottomrule
\end{longtable}

%% file: sections/results/scaffold/side-chain.tex
\subsubsection{Analyzing the structural distribution of side chains}

\begin{figure}[t!]
\centering
\includegraphics{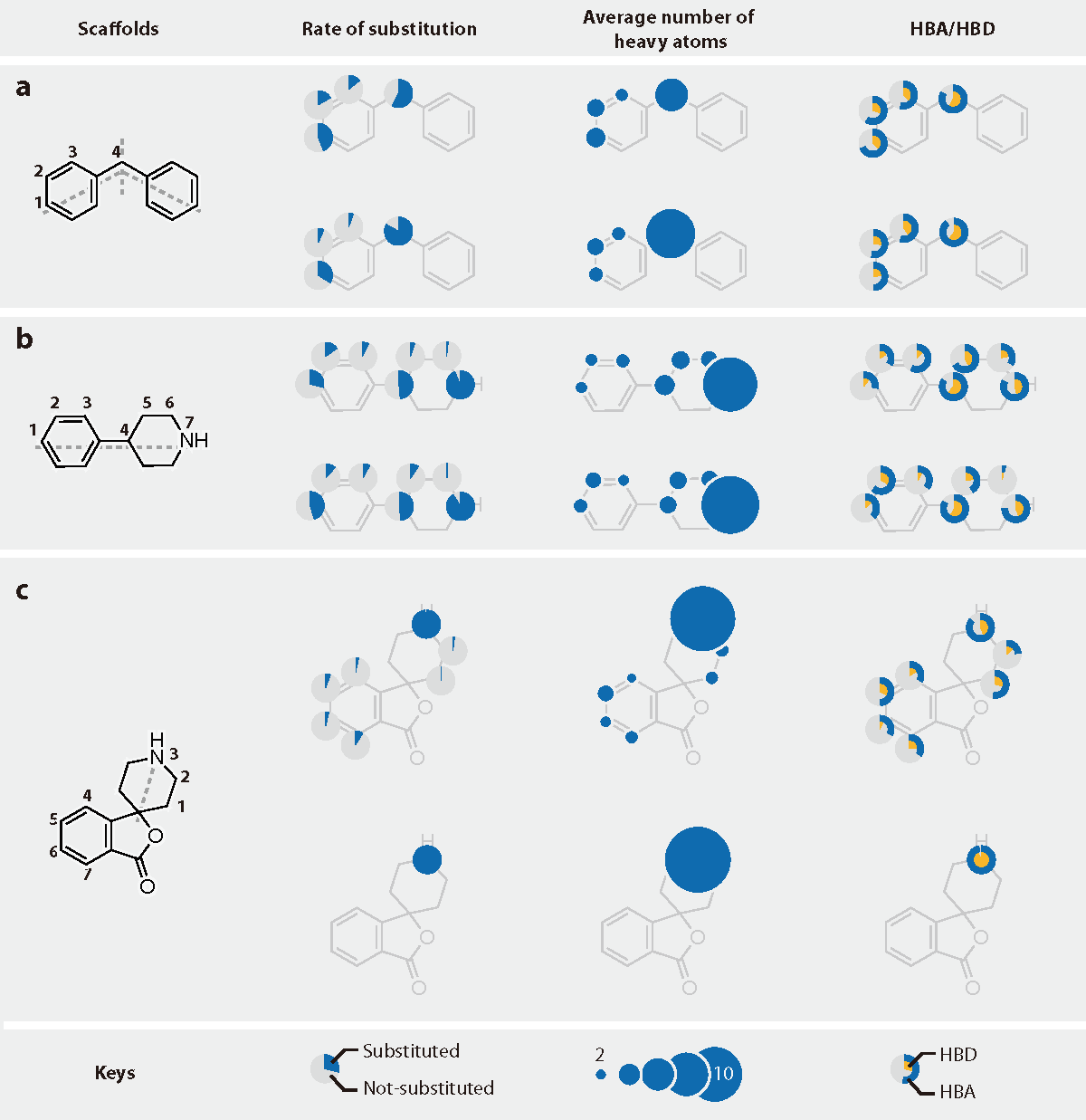}
\caption{
    The distribution of side chain properties at each location for scaffold
    \textbf{12}-\textbf{14}.
}
\label{fig:side-chain}
\end{figure}

Finally, we investigate the structural distribution of side chains
generated by the model. We summarize the result in
\ref{fig:side-chain}.

The result demonstrated a diversified substitution pattern generated
from the model. Note that for scaffold \textbf{14}, the
substitution only happens in location \textbf{3} for molecules
inside ChEMBL (the reason for the low diversity is discussed in
Section \ref{sec:bad-case}). But in the generated molecules, the substitution pattern
is much more diverse, demonstrating high generalizability for the model.

Interestingly, it is found that the model adds side-chain to nitrogen
atoms more often than other locations. It is also fond the size of
side-chain attached to nitrogen atoms tends to be much larger compared
with that in other positions. Those observations could be result from
a mixed effect of synthetic accessibility and drug-likeness.

%% file: sections/results/scaffold/docking.tex
\subsubsection{Distribution of docking scores}

\begin{figure}[t!]
\centering
\includegraphics{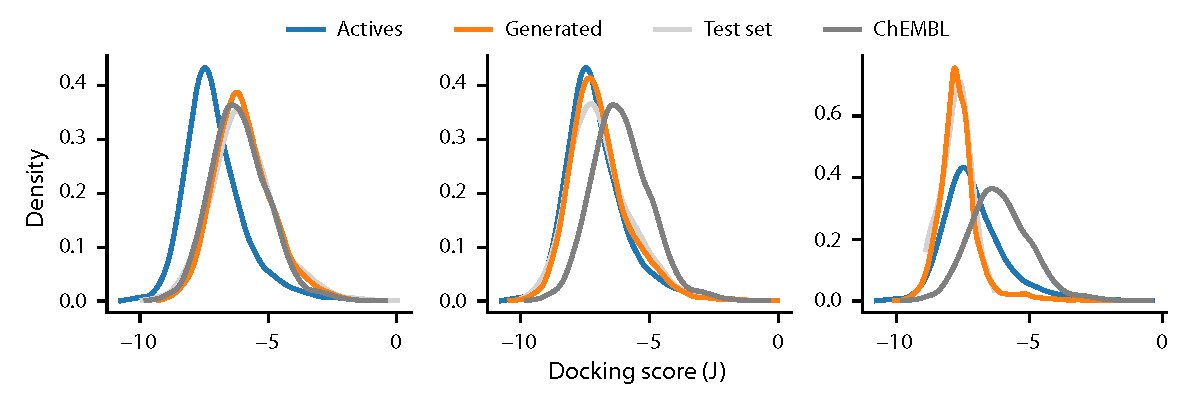}
\caption{
    The distribution of docking scores.
    This
    plot shows the distribution of docking scores of three sets of molecules
    (active molecules, generated molecules, test set molecules and random
    compounds from ChEMBL) for each scaffold (\textbf{a.} scaffold
    \textbf{12}, \textbf{b.} scaffold
    \textbf{13} and \textbf{c.} scaffold
    \textbf{14}).
}
\label{fig:docking}
\end{figure}

The distribution of docking scores of generated molecules with DRD2 is
shown in Figure \ref{fig:docking}. As demonstrated, the predicted activities
of generated samples (orange) are similar to that of the test set
molecules (light grey). For scaffold \textbf{13} and
scaffold \textbf{14}, the predicted activity against DRD2
for generated samples are significantly higher compared with randomly
sampled molecules from ChEMBL, showing that the model is good at
utilizing privileged structures for generating bioactive molecules. On
the other hand, for scaffold \textbf{12}, the predicted
activities of generated samples could not outperform that of random
ChEMBL samples. This may comes from the fact that the bioactivity privilege
of scaffold \textbf{12} is not as significant compared
with \textbf{13} and \textbf{14},
according to the docking result. In fact, scaffold
\textbf{12} is highly prevalent among molecules in
ChEMBL, and could occur in compounds with different targets. This may
explain the results for scaffold \textbf{12} shown above.

%% file: sections/conclusion.tex
\section{Conclusion}

We proposed a comprehensive solution for scaffold-based drug discovery using
deep learning. The model is capable of performing molecular design tasks based on a wide spectrum of 
scaffold definitions, including BM-scaffolds, cyclic skeletons, as well as 
scaffolds with specifications on side-chain properties. Evaluation is performed
to access the performance of the models. In addition, we choose several  privileged scaffolds
for GPCRs for the case study, demonstrating the practical value of the model in
medicinal chemical problems. We intend to make this tool freely available for 
academical purposes. Please contact the authors for licensing information.